# Size-Dependent Structural Phase Transitions in SrTiO$_3$ Nanoparticles

Han Zhang[1], Sizhan Liu[1], Megan E. Scofield[2], Stanislaus S. Wong[2,3], Xinguo Hong[4], Vitali Prakapenka[5], Eran Greenberg[5] and Trevor A. Tyson[1,*]

[1]Department of Physics, New Jersey Institute of Technology, Newark, NJ 07102
[2]Department of Chemistry, State University of New York at Stony Brook, Stony Brook, NY 11794-3400
[3]Condensed Matter Physics and Materials Science Division, Brookhaven National Laboratory, Upton, NY 11973
[4]Mineral Physics Institute, Stony Brook University, Stony Brook, NY 11794
[5]Center for Advanced Radiation Sources, University of Chicago, Argonne, Illinois 60439, USA

*Corresponding Author:  Trevor A. Tyson, e-mail: tyson@njit.edu

## Abstract

Understanding the structural phase diagram of nano scale SrTiO$_3$ has important implications on the basic physics and applications of the general class of transition metal oxide perovskites.  Pressure dependent structural measurements on monodispersed nanoscale SrTiO$_3$ samples with average diameters of 10 to ~80 nm were conducted.  A robust pressure independent polar structure was detected in the 10 nm sample for pressures of up to 13 GPa while a size dependent cubic to tetragonal transition occurs (at $P = P_c$) for larger particle sizes.  The results suggest that the growth of ~10 nm STO particles on substrates with large lattice mismatch will not alter the polar state of the system for a large range of strain values, possibly enabling device use.

PACS Number(s): 77.80.-e, 64.60.-i, 61.46.-w, 62.50.-p

Perovskites form a particularly interesting class of materials, because even slight modifications of the crystal structure can lead to drastic changes in physical properties, such as superconductivity, ferroelectricity, and ferromagnetism. Strontium titanate, $SrTiO_3$ (STO), is generally considered to be a model perovskite and plays an important role in the understanding of soft-mode-driven phase transitions, which have been extensively studied for more than five decades.

$SrTiO_3$ is known as an incipient ferroelectric material, with the pure form remaining paraelectric down to ~0 K. Small perturbations of the structure, such as but not limited to isotope substitution [1], chemical doping [2], the application of electric fields [3], and/or stress [4], can result in the onset of a ferroelectric state at finite temperatures. For example, isotopic substitution with $^{18}O$ yields evidence of a ferroelectric transition temperature with $T_c$ ~ 23 K. The largest enhancement of $T_c$ was first reported by Haeni *et al.* When applying tensile strain in STO epitaxial films, a ferroelectric transition can be observed above room temperature [5]. It has also been reported that the phase transition behavior and domain structure in anisotropically-strained STO thin films differ significantly as compared with isotopically-strained films [6]. In the asymmetrically-strained STO system, the anisotropy of properties along two different directions is observed in measurements of switchable polarization, relaxor character, and permittivity. According to a study conducted by Jang *et al*., the role of strain is to stabilize longer range correlations of pre-existing nanopolar regions (PNRs) [7], related to unintentional Sr deficiency.

The bulk STO system is cubic under ambient conditions. However, the oxygen atoms located in the octahedral sites can easily rotate around the central titanium atom, thereby giving rise to possible distortions of the cubic crystal [8]. An antiferrodistortive (AFD) transition to a non-polar tetragonal phase occurs at ~ 105 K at ambient pressure. This temperature-induced phase transition has been intensively examined [9], both experimentally and theoretically, with a soft phonon mode model proposed to describe the observed transition.

The AFD transition of the STO system can also be affected by applying pressure. In earlier experimental work on bulk samples, no sign of any anomaly in the elastic constants was observed between 0.4 to 2.2 GPa [10]. The transition temperature was found to have shifted linearly with increasing pressure below 0.85 GPa, and then altered nonlinearly above that pressure [11]. Later, evidence of the AFD transition induced by pressure was observed through a combination of experiments including Brillouin scattering (BS) [12], high pressure Raman measurements [13], and X-ray absorption fine structure (XAFS) spectroscopy [14]. Although the pressure associated with the onset of the transition differed with each independent experimental method, the pressure values previously investigated fell within a range of 5.5 – 7.5 GPa. A theory explaining the phase transition within the STO system was also developed using *ab initio* density functional theory (DFT) methods, which predicted a transition near 6 GPa [15]. However, more recent experimental work has tended to favor a higher pressure for the detection of such a phase transition. In effect, Guennou *et al*. [16] utilized single crystal X-ray diffraction (XRD) and Raman spectroscopy methods to report on a transition pressure of 9.6 GPa, a finding which is also further supported by recent work [17].

Studies of simple nanoscale perovskites have been conducted, with the aim of utilizing them in device applications for high density storage. For example, the ferroelectric behavior of $BaTiO_3$ nanoparticles (see Ref. [18]) has been shown to disappear if the particle size is below a critical size of 5–10 nm. Meanwhile, in the STO system, lattice expansion [19] and nano-size effects [20] have been observed with reduced particle size. The transition temperature from the classical paraelectric to the quantum paraelectric state has been reported to increase with decreasing particle size, which contrasts with the typical trend expected between transition temperature and size effects. For example, the paraelectric to ferroelectric phase transition found in $BaTiO_3$ and $PbTiO_3$ samples experienced a decrease in transition temperature with a corresponding decrease in particle size [21].

In our recent work conducted on various sizes of STO nanoparticles, a polar state over a wide temperature range (possibly with ferroelectric properties) was found in free-standing 10 nm nanoparticles [22]. More recently, a scaled reduction of STO thin films was found to have produced ferroelectricity [23]. We note that while a large body of research has previously been conducted on STO films, very limited studies on ferroelectricity and on the polar state of STO nanoparticles have been carried out. Nanoscale STO can potentially possess very unique properties that are not exhibited by either bulk or film systems. It is very likely that nano-size STO particles will exhibit distinctly different properties for bulk samples.

To assess this system more thoroughly, a systematic study of STO particles possessing various particle sizes is needed. In this work therefore, the phase diagram of STO nanoparticles will be constructed by analyzing and studying samples, incorporating a broad range of average particle diameters, under varying pressure conditions. Particles averaging 20 nm to ~83 nm in diameter are found to exhibit size-dependent structural phase transitions, which occur at a lower pressures with reduced particle size. Conversely, the 10 nm particles exhibit a robust pressure-independent polar structure forpressures of up to 13 GPa.

Our monodispersed nanoparticles of $SrTiO_3$ are synthesized by soft chemistry methods. (See supplementary document) High-resolution synchrotron x-ray diffraction as well as x-ray spectroscopic and Raman methods reveal that the samples are not only stoichiometric and monodispersed but also exhibit a polar structure for the 10 nm particle size [22]. Studies of Wu *et al.* [19] indicate that polar modes were present in all nanoscale STO particles < 83 nm in diameter, and the polar states were enhanced with reducing particle size. In this work, particle diameters and morphologies were assessed by transmission electron microscopy (TEM) measurements [22] with the various STO samples possessing average diameters of 10.1 ± 1.1 nm, 20 ± 2.0 nm, 40 ± 4.0 nm, and 82.6 ± 9.1 nm, respectively.

High pressure X-ray diffraction experiments on samples possessing diameters of 10 nm and 83 nm were conducted at the X17C beamline at the National Synchrotron Light Source (NSLS) at Brookhaven National Laboratory (BNL). A focused beam of dimension 20 μm × 23 μm and a wavelength of 0.409929 Å were used. The sample-detector distance was 287.1 mm. The diffraction patterns were collected with a Rayonix 165 charge coupled device (CCD) detector. All experiments were performed in a diamond-anvil cell (DAC) with stainless steel gaskets and with a 4:1 methanol-ethanol mixture as the pressure transmitting medium. Several ruby chips were placed in different parts of the chamber for pressure measurements based on the ruby fluorescence wavelengths, and the average reading from three different rubies was used to determine the pressure. Four 4000-second scans were collected for each pressure for the 83 nm sample, while two 4000-second scans were collected for the 10 nm sample for each pressure. The 2θ range collected was up to 23º for both data sets.

Experiments performed on the STO particles possessing average diameters of 20 nm and 40 nm were conducted at beamline 13-ID-D at the Advanced Photon Source (APS) at Argonne National Laboratory (ANL). Measurements were conducted with a wavelength of 0.3344 Å and a beam of dimensions 4 μm ×3 μm was used. The sample-detector distance was 194.2 mm. A 240 μm hole in a rhenium gasket, pre-indented to 50 μm, served as a sample chamber. Ne gas was used as the pressure transmitting medium and gold particles were placed in the DAC for the pressure measurement. At each pressure, scans lasting 60 s in duration were collected and processed. The 2θ range was 0 to 24° for the 20 nm sample, and 0 to 12° for the 40 nm sample, respectively. The variations in range are a reflection of the corresponding differences in the opening angles of the seats of the different DACs used in the measurements. A sample raw data pattern is given at Fig. S1 in the Supplementary Information section.

The programs Fit2D [24] and Dioptas [25] were utilized to integrate the two dimensional diffraction images to yield the one dimensional intensity *vs.* 2θ XRD patterns. Rietveld refinements on the XRD data were conducted by using the JANA2006 software package [26].

Fig. 1 highlights examples of intensity *vs.* 2θ XRD patterns, taken at each beamline mentioned above (see also Fig. S1 in the Supplementary Information section). The transformation into 2θ-space revealed that the peaks from the pattern of the 83 nm sample illustrated in Fig. 1(a) are sharper than those from the 20 nm sample, presented in Fig. 1(b), mainly due to the difference in particle size [27]. The signal from the Ne pressure medium was marked as * in Fig. 1(b). With the exception of that signal, no new peaks appear over the whole measured pressure range. Additionally, expected peaks appear to shift to higher angles with increasing pressure, as seen in both Fig. 1(a) and 1(b). Rietveld refinements were performed on the collected XRD data to obtain the structural parameters of the STO particles. The profiles of the refinement (data fit) under ambient pressure for the 83 nm and 20 nm sample are shown in Fig. S2(a) and Fig. S2(b) in the Supplementary Information section. It should be noted that the experimental setup for both samples were different from one another (see experimental details), which sometimes led to the same Bragg peak appearing at two different $2\theta$ angles.

In order to better understand the correlation between pressure and particle size, a first order equation of state fit using the Murnaghan equation was performed with the results presented in Fig. 2. Fig. 2(b) highlights the fitting curve when using the same Murnaghan equation through the entire pressure range, with a clear deviation observed between 6 and 10 GPa, indicated as $P_c$. Meanwhile, fitting the experimental results in two different regions, as shown in Fig. 2(a), gives significantly improved refinement results, indicating a non-continuous compression process [28] with a structural phase transition at ~ 6 GPa. Meanwhile, the bulk modulus $B_0$ and its pressure derivative $B_0'$ obtained from the fit are $B_0 = 153.23 \pm 9.48$ GPa and $B_0' = 10.35 \pm 2.28$ GPa, respectively. $B_0'$ is slightly off the normal range, with typical values corresponding to $B_0'$ between 2 and 8 [29]. The value of $B_0$ can be compared with the corresponding value of other well-studied perovskites, which are listed in Table S1 in the Supplementary Information section. It can be observed that although the 83 nm sample possesses a structural transition near 6 GPa, which is similar to the bulk value, the bulk modulus is significantly smaller than that of the bulk sample (~220 GPa). This result can be caused by the size effects associated with the nanoscale particles. The expansion of the STO lattice with reduced size suggests a lattice softening with size reduction, consistent with the $B_0$ reduction. The expansion is due to the reduction of the hybridization of the oxygen $p$ and titanium $d$ bands (see previous work in Ref. [22] supplementary text). This behavior is also observed in Ref. [30], in which the decrease in the electrostatic force caused by the valence reduction of Ce ions and an increase in the ionicity of Ti ions were argued to be the reason for the observed lattice expansions in $CeO_{2-x}$ and $BaTiO_3$ nanoparticles, respectively. Similar analysis was not conducted with the 40 nm sample, due to the lower 2θ range with that data set, as mentioned in the experimental details. However, by investigating the data associated with the 20 nm sample, as shown in Fig. 2(c), a deviation (transition) is observed but is shifted towards a lower pressure value of ~2.5 GPa. Additionally, the observed modulus value was found to have decreased as compared with typical bulk values.

Fig. 2(d) presents the first order equation–of-state fit for the 10 nm STO sample. The experimental data can be well described with a single first order equation in the whole pressure range, which is quite different from the 83 nm sample. The good agreement of the data with the fitting curve suggests no structural transition in this pressure range. The pressure derivative $B_0^{'}$ extracted from the refinement is ~15, which is significantly larger than the normal value. This enhancement is typical of anisotropic compression [31]. This behavior is in good agreement with our previously reported work [22] that demonstrated that the 10 nm STO sample is polar over a wide temperature range.

In order to view the pressure-dependent structural change, we further investigated the XRD pattern by examining the change in peak widths as a function of pressure (looking for splittings/broadening caused by structural transitions) [32]. The results are presented in Figure 3. The basic idea is that when the sample undergoes a transition from the cubic to the tetragonal phase, some of the specific peaks in the XRD pattern will split. Although the splitting can be small under many circumstances, it can be well detected when studying the peak width *vs.* the pressure.

In this work, we first fit some chosen peaks with a Lorentzian profile, and then retrieved the corresponding peak width *vs.* pressure curve. The original data can be found in the Supplementary Information section. To visualize the changes clearly, we normalized the peak width as $\varpi_{nor} = \frac{\varpi/\varpi_0}{\varpi^{(101)}/\varpi_0^{(101)}}$, in which $\varpi_{nor}$ is the normalized peak width; $\varpi$ is the original peak width; $\varpi_0$ is the original peak width at lowest pressure; $\varpi^{(101)}$ is the width of peak (101), which will not spilt during the cubic-tetragonal transition, and $\varpi_0^{(101)}$ is the width of peak (101) at lowest pressure. The results are displayed in Fig. 3. Fig. 3(a) includes the peak width *vs.* pressure data for the (112), (002), (111), and (101) peaks, which can be attributed to the 10 nm STO sample. The 10 nm sample yields no obvious trends which can be observed through the entire pressure range of all of these four peaks. The (112) and (002) peaks exhibit an almost linear response between ambient pressure and 13 GPa, suggesting that the sample structure is stable in this pressure range. However, as we increased the STO particle size, there are clear deviations in the graphs (Fig. 3(b), 3(c) and 3(d)). For example, for the 20 nm sample, all four peaks maintain almost the same width up to ~2.5 GPa, but a sudden increase is observed near ~2.5 GPa in the (112) and (002) peaks, which is likely related to the structural transition from the cubic to the tetragonal phase [32]. This finding serves as strong evidence that the 20 nm STO has given rise to a pressure-induced phase transition at ~2.5 GPa. As the particle size is increased, the transition pressure is seen to shift towards the higher pressure region. In Fig. 3(c), the transition pressure is found to be ~4.5 GPa for the 40 nm sample, and in Fig. 3(d), the pressure is noted to be ~6.0 GPa for the 83 nm sample.

We further investigated this transition pressure as a function of different particle sizes, based on the theory developed by Chen et al. [33], correlating pressure-induced changes with size-induced changes in the transition temperature. The relationship between the transition pressure $P_{cj}$ and the particle size $K_j$ is found to be related by the following expression: $P_{cj} \sim P_{cb}\left(A - \frac{B}{K_j}\right)$, where $P_{cb}$ is the transition pressure of bulk sample and A and B are constants determined by the nature of the chemical bonding inside the sample.

The theory suggests a linear correlation between the transition pressure and the inverse of the particle size. Examination of Fig. 4(b) reveals that this prediction holds for this system. These combined results are found in Fig. 3 and Fig. 4 and give corroborative evidence for our previous conclusion that the 20 nm, 40 nm, and 83 nm STO samples exhibit a pressure-induced phase transition whose onset appears to decrease with decreasing particle size, whereas the 10 nm STO is stable (in a polar phase) between ambient pressure and 13 GPa. The structural phase diagram under pressure is presented as Fig. 4(a).

This result provides insight into the possible application of STO nanoparticles of various sizes in data storage devices. Polar nano-scale STO particles can be synthesized from simple wet chemical methods. One can grow or deposit the nanoparticles (of diameter ~ 10 nm) onto substrates with significant lattice mis-match without appreciably altering the polar state. The application of an electric field to orient the particles followed by annealing may possibly produce a high density nanoscale array of nanoferroelectric materials. By controllably depositing the particles onto a densely patterned surface, high capacity storage may be enabled.

The bulk phase of $SrTiO_3$ (STO) is paraelectric and exhibits a structural phase transition near ~6 GPa under hydrostatic pressure. We have conducted pressure-dependent structural measurements (up to ~20 GPa) of monodispersed nanoscale samples with average diameters of 10 nm, 20 nm, 40 nm, and 83 nm, respectively. The transition pressure was found to have decreased with decreasing particle size, and a robust pressure-independent structure of the 10 nm sample was observed. The results suggest that growth of STO nanoparticles onto substrates which do not match the underlying STO lattice will not alter the polar state of the system for a large range of strain values, thereby enabling more widespread device use.

This work is supported by the U.S. Department of Energy (DOE) Grant DE-FG02-07ER46402. Research support for MES and SSW was provided by the U.S. Department of Energy, Basic Energy Sciences, Materials Sciences and Engineering Division. Synchrotron powder x-ray diffraction experiments were performed at Brookhaven National Laboratory's National Synchrotron Light Source (NSLS) and Advanced Photon Source (APS). Use of the NSLS was supported by the U.S. Department of Energy, Office of Science, Office of Basic Energy Sciences, under Contract No. DE-AC02-98CH10886. This research used resources of the Advanced Photon Source, a U.S. Department of Energy Office of

**Supporting Information.** Synthesis of $SrTiO_3$ nanoparticles, high resolution PDF image, Rietveld refinement results, Table of bulk modulus and normalized peak width *vs.* pressure.

# Figure Captions

**Fig. 1.** High pressure synchrotron XRD patterns of SrTiO$_3$ with average particle sizes: (a) 83 nm and (b) 20 nm. The 2θ values of (b) have been re-calculated for easy comparison. Peak positions shift systematically to higher 2θ values with increasing pressure. Note that the experimental setups for the two STO samples were different (wavelength and sample detector distances). Therefore the same Bragg peaks in the two sets of data appear at different 2θ angles.

**Fig. 2.** (a) The first order Murnaghan equation of state fit (solid lines) of experimental results (circles) for 83 nm STO. The fitting was conducted through two different regions (blue and green circles, respectively). (b) The first order Murnaghan equation of state fit (solid line) of experimental results (circles) in the entire data region, with a deviation that can be observed between 6 and 8 GPa. (c) Single model fit over the entire pressure range for the 20 nm sample. (d) Single model fit over the entire pressure range for the 10 nm sample.

**Fig. 3.** Specified peak widths vs. pressure for the STO system extracted from the diffraction patterns for particle sizes (a) 10 nm (b) 20 nm, (c) 40 nm, and (d) 83 nm, respectively. There is a clear shift of the pressure towards the lower pressure region with decreasing particle size. The widths have been normalized as $\varpi_{nor} = \dfrac{\varpi/\varpi_0}{\varpi^{(101)}/\varpi_0^{(101)}}$.

**Fig. 4.** Phase diagram of SrTiO$_3$ nano particles, with transition pressure *vs.* (a) particle size and (b) the inverse of the particle size. The solid lines are refined, based on the results in ref [33].

Fig. 1. Zhang *et al.*

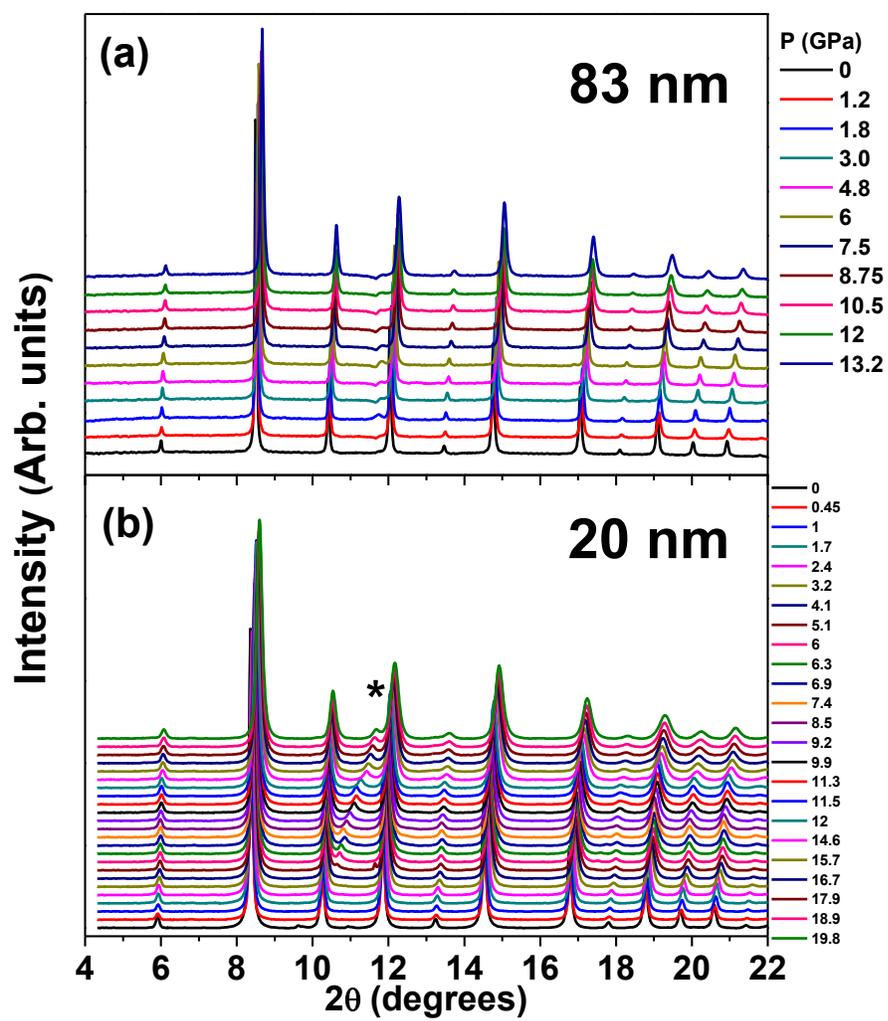

**Fig. 2.** Zhang *et al.*

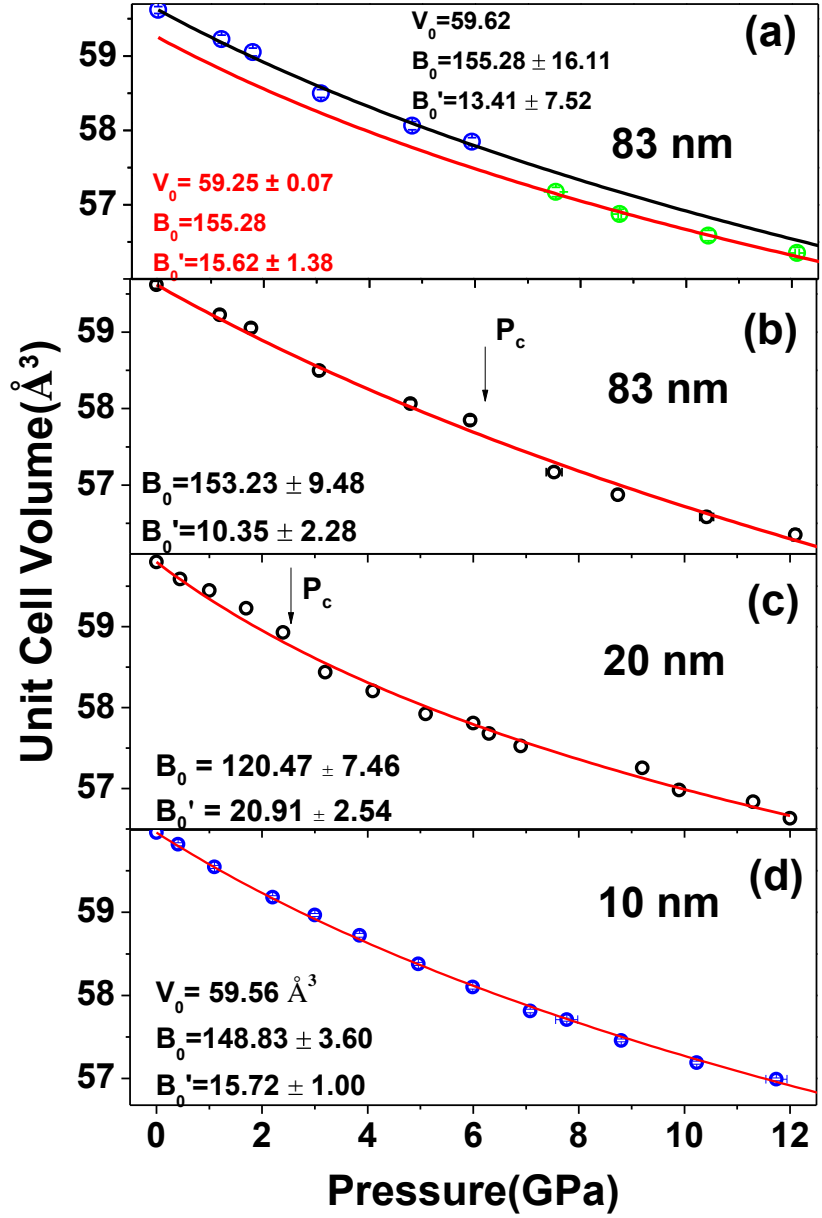

**Fig. 3.** Zhang *et al.*

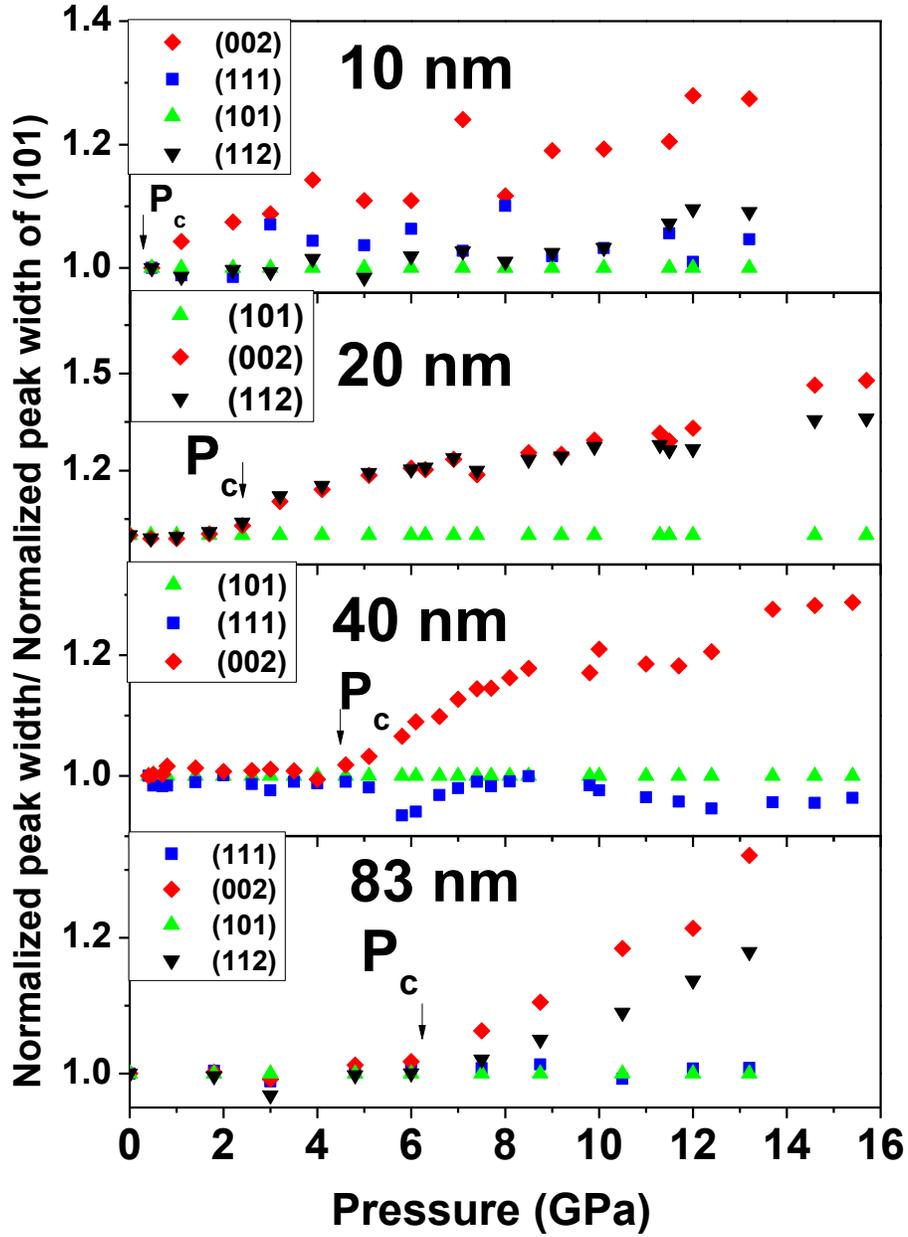

**Fig. 4.** Zhang *et al.*

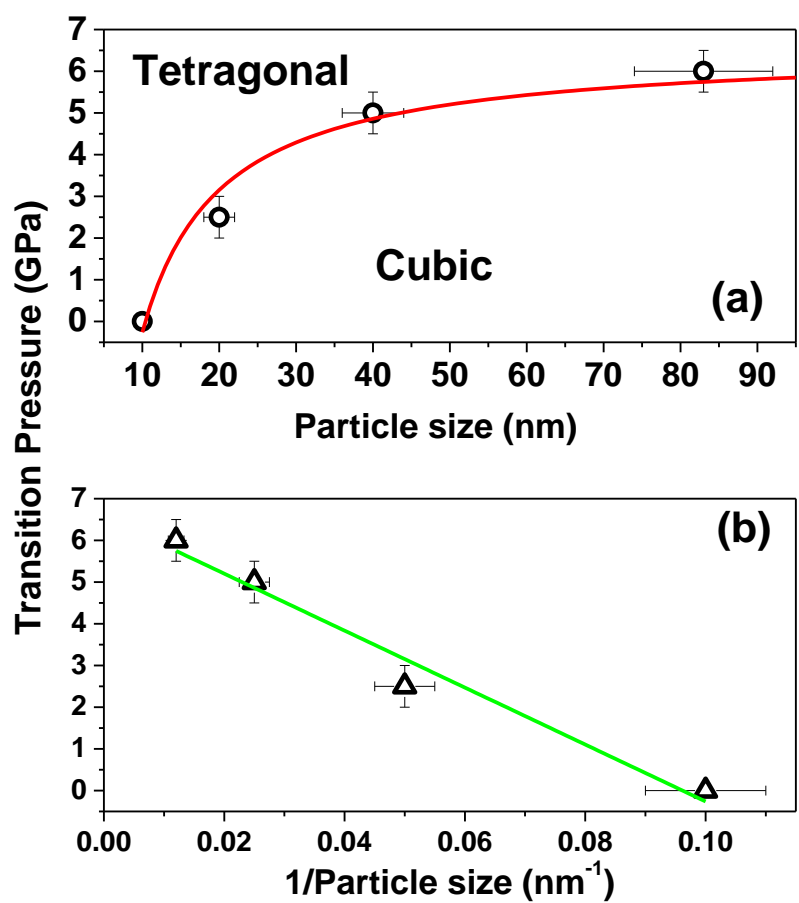

# Size-Dependent Structural Phase Transitions in SrTiO$_3$ Nanoparticles
# (Supplementary Information)

Han Zhang[1], Sizhan Liu[1], Megan E. Scofield[2], Stanislaus S. Wong[2,3], Xinguo Hong[4], Vitali Prakapenka[5], Eran Greenberg[5], and Trevor A. Tyson[1,*]

[1]Department of Physics, New Jersey Institute of Technology, Newark, NJ 07102
[2]Department of Chemistry, State University of New York at Stony Brook, Stony Brook, NY 11794-3400
[3]Condensed Matter Physics and Materials Science Division, Brookhaven National Laboratory, Upton, NY 11973
[4]Mineral Physics Institute, Stony Brook University, Stony Brook, NY 11794
[5]Center for Advanced Radiation Sources, University of Chicago, Argonne, Illinois 60439, USA

*Corresponding Author:
Trevor. A Tyson, e-mail: tyson@njit.edu

## I. Experimental Methods

Our monodisperse nanoparticles of SrTiO$_3$ are synthesized by soft chemistry methods. 10 nm SrTiO$_3$ nanoparticles were prepared by employing a hydrothermal technique [1]. More specifically, solutions of titanium bis(ammonium lactate) dihydroxide (TALH) and strontium hydroxide octahydrate were mixed together in a 1: 1 molar ratio with the pH adjusted to 13.5. Hydrazine and oleic acid were added to an autoclave with the precursor solution and heated to 120°C for 24 hrs. The sample was then isolated upon centrifugation and washed with water followed by ethanol.

The preparation of 20 nm $SrTiO_3$ nanoparticles was carried out by following the protocol of ref. [2], adding 0.006 mol of $Ti(OC_4H9)_4$ and 0.0072 mol of $Sr(OH)_2$ x $8H_2O$ to n-butylamine (20 mL), and the mixture was allowed to stir for 2 hrs. Subsequently, it was transferred to a Teflon-lined autoclave and run at 180°C for 24 hrs. The resulting white precipitate was isolated and initially washed with diluted formic acid, followed by aliquots of deionized water and ethanol. The washing process was subsequently repeated 3 times each. The $SrTiO_3$ sample was then allowed to dry overnight at 80ºC.

40 nm $SrTiO_3$ nanoparticles were synthesized using a hydrothermal protocol [3]. Specifically, $TiO_2$ powder (0.18 g, 2.3 mmol) was mixed in a 20 mL aqueous solution of KOH (1.26 g, 23 mmol) to which $Sr(OH)_2·8H_2O$ (0.508 g, 2.3 mmol) was also added. The mixture was placed within a 23 mL autoclave and heated to 150°C for 72 hrs. Once the autoclave had cooled to room temperature, the resulting product was isolated, washed with water, and subsequently allowed to dry overnight.

The 83 nm diameter $SrTiO_3$ nanoparticles were generated by methods described in reference [4]. The sample was prepared by grinding powders of strontium oxalate, anatase titanium dioxide, and sodium chloride (1: 1: 20 molar ratio) with a mortar and pestle for 30 min, respectively, until the mixture became homogeneous. Nonylphenyl ether (NP-9 with a molar ratio of 3) was then added to the mixture and further ground until the mixture was also uniform and homogeneous. The combination was then loaded into a porcelain reaction boat and heated in a tube furnace for 3.5 hrs at 820°C. The resulting powder was isolated after centrifugation, and subsequently washed with water followed by ethanol to remove any excess salt.

## II. Results

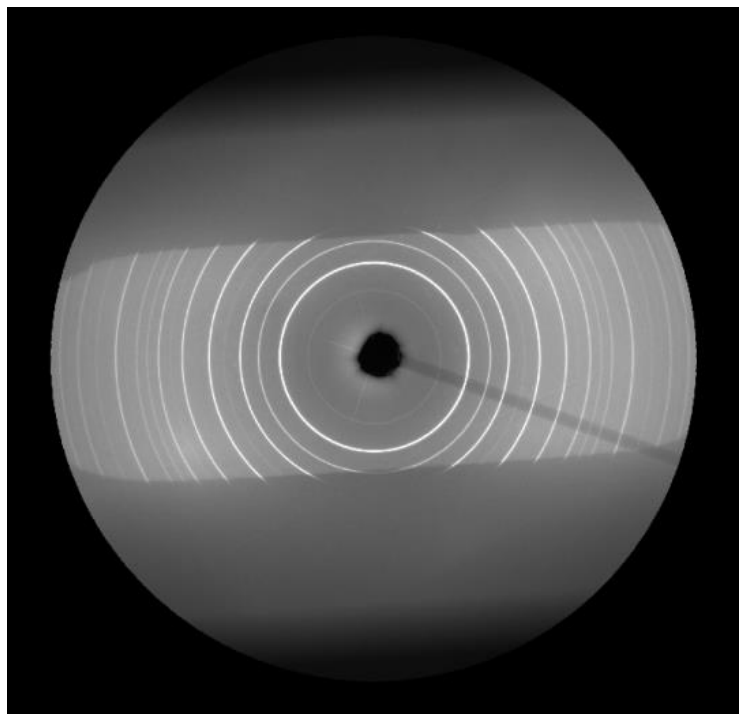

**Fig. S1** Two-dimensional high pressure XRD data pattern taken at APS 13-ID-D with a CCD detector. The data collected is of high quality with no bright spots from single crystal particles.

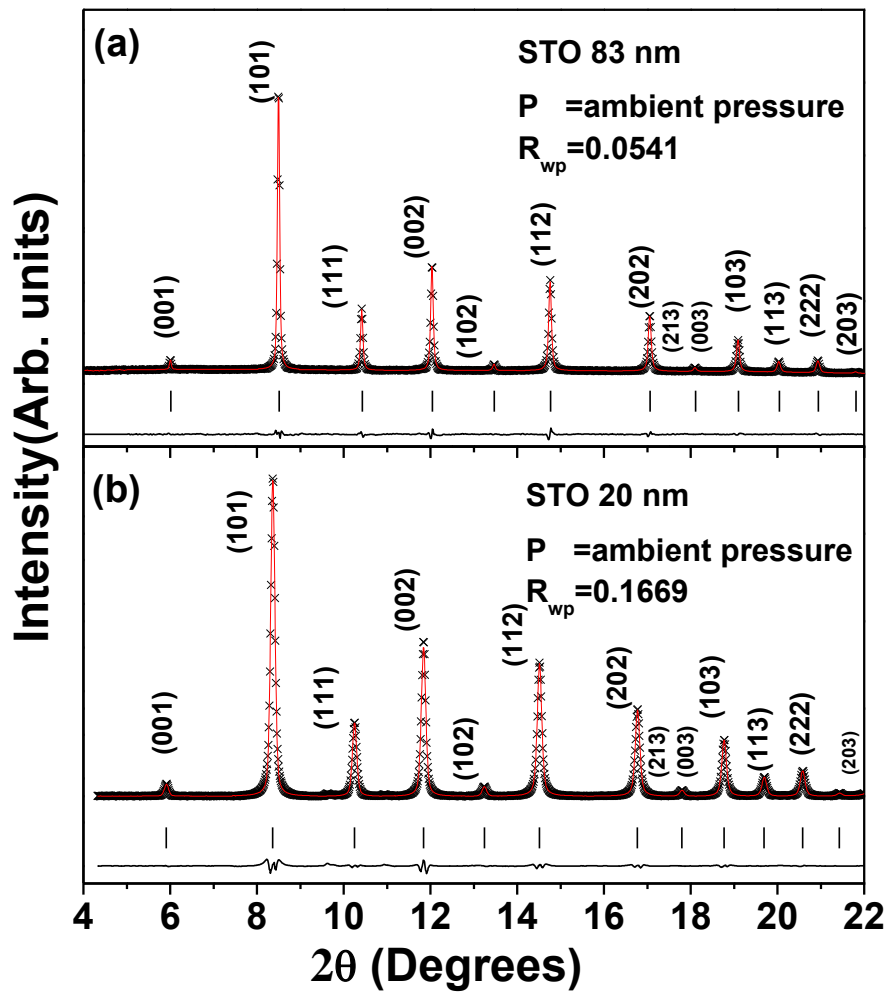

**Fig. S2.** Rietveld refinement results at ambient pressure for (a) 83 nm and (b) 20 nm STO. The observed (crosses), calculated (solid line), and difference (bottom line) patterns are shown. The vertical bars show the peak positions for the refined model. The 2θ values of (b) have been re-calculated for easy comparison.

**Table S1. Bulk Modulus Values**

| Sample | B (GPa) | B' |
|---|---|---|
| $PrTiO_3$ (cubic) | 195(3) [5] | 4 |
|  | 237(4) [6] | 4.5 |
|  | 141(5) [7] | 4 |
| $PrTiO_3$ (Tetragonal) | 100(3) [5] | 4 |
|  | 107(3) [6] | 5 |
|  | 104(4) [7] | 4 |
| $BaTiO_3$ (cubic) | 162 [8] |  |
| (Tetragonal) | 141 [9] |  |
| $EuTiO_3$ (cubic) | 180.1 [10] |  |
| (Tetragonal) | 190.3(4) [10] |  |
| $SrTiO_3$ (cubic) | 176(3) [11] | 4.4 (8) |
| (Tetragonal) | 225(6)* [12] | 4.7 (4) |
| $SrTiO_3$ nano-particles (This Work) |  |  |
| (10 nm) | 148.83 | 15.72 |
| (20 nm) | 120.47 | 20.91 |
| (83 nm) | 153.23 | 10.35 |

* Calculated with a third-order Birch-Murnaghan Equation of State

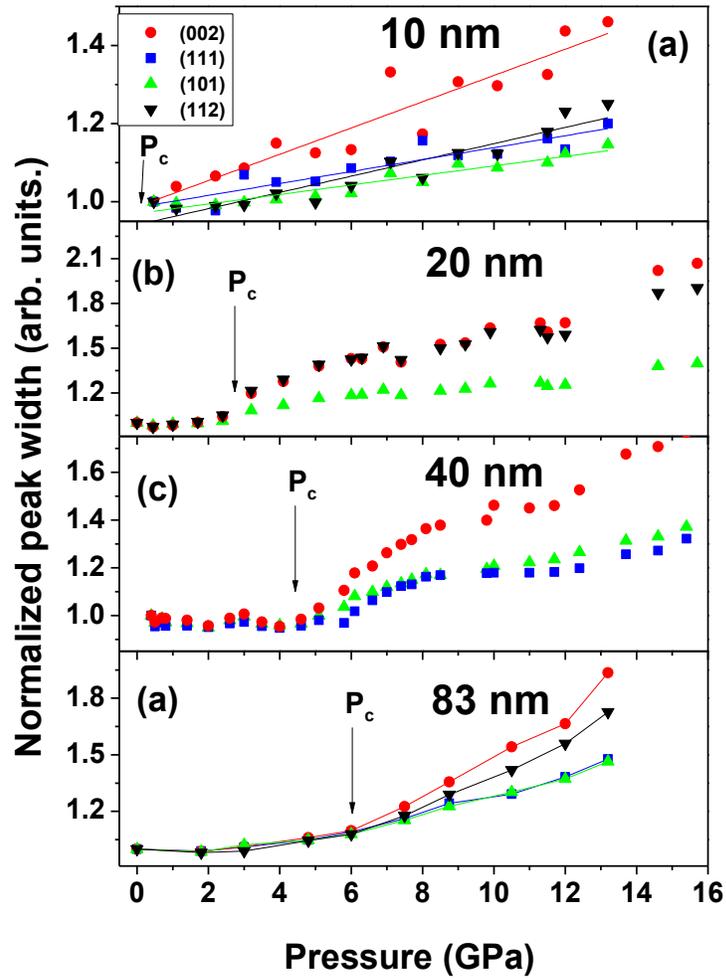

**Fig. S3.** Specified peak widths vs. pressure for the STO system extracted from the diffraction patterns for particle sizes (a) 10 nm (b) 20 nm, (c) 40 nm, and (d) 83 nm, respectively. There is a clear shift of the pressure towards the lower pressure region with decreasing particle size.